\documentclass[aps,prl,superscriptaddress,reprint]{revtex4-1}
\usepackage{graphicx}
\usepackage{color}
\usepackage{bm}
\usepackage{amsmath,amssymb}
\newcommand{\be}{\begin{equation}}
\newcommand{\ee}{\end{equation}}
\newcommand{\ba}{\begin{array}}
\newcommand{\ea}{\end{array}}
\newcommand{\bqa}{\begin{eqnarray}}
\newcommand{\eqa}{\end{eqnarray}}

\begin{document}


\title{Could the $X(3915)$ and the $X(3930)$ Be the Same Tensor State?}


\author{Zhi-Yong Zhou}
\email[]{zhouzhy@seu.edu.cn}
\affiliation{Department of Physics, Southeast University, Nanjing 211189,
People's Republic of China}
\affiliation{State Key Laboratory of Theoretical Physics, Institute of Theoretical Physics, Chinese Academy of Sciences, Beijing 100190, China}

\author{Zhiguang Xiao}
\email[]{xiaozg@ustc.edu.cn}
\affiliation{Interdisciplinary Center for Theoretical Study, University of Science
and Technology of China, Hefei, Anhui 230026, China}
\affiliation{State Key Laboratory of Theoretical Physics, Institute of Theoretical Physics, Chinese Academy of Sciences, Beijing 100190, China}

\author{Hai-Qing Zhou}
\email[]{zhouhq@seu.edu.cn}
\affiliation{Department of Physics, Southeast University, Nanjing 211189,
People's Republic of China}
\affiliation{State Key Laboratory of Theoretical Physics, Institute of Theoretical Physics, Chinese Academy of Sciences, Beijing 100190, China}

\date{\today}

\begin{abstract}

By using a combined amplitude analysis of the $\gamma\gamma\rightarrow
D\bar{D}$ and $\gamma\gamma\rightarrow J/\psi\omega$ data,
we demonstrate that the $X(3915)$, which is quoted as a $J^{PC}=0^{++}$ state in
the Particle Data Group table, is favored by the data to be a $J^{PC}=2^{++}$ state
appearing in both channels, which means that  the $X(3915)$ and the
$X(3930)$ can be regarded as the same $J^{PC}=2^{++}$ state.
Meanwhile, the data also prefer a large helicity-0 contribution of
this tensor resonance to the amplitudes  instead of the helicity-2
dominance assumed by {\it BABAR}, which may indicate a sizable
portion of non-$q\bar q$ components in this state.
Identifying
the $X(3915)$ with the $X(3930)$   and abandoning the helicity-2 dominance
for this tensor state are helpful for the further understandings of the
properties of this state and also of the mysterious ``$XYZ$''
charmoniumlike resonances.

\end{abstract}
\pacs{14.40.Pq, 13.75.Lb, 11.80.Et}

\maketitle

In recent years, more than a dozen 
new charmoniumlike ``$XYZ$'' states above open-charm thresholds have
been observed from experiments~\cite{Agashe:2014kda}.  However,
these newly observed states
seem to deviate from the predictions of the quark potential
models~(see Ref.~\cite{Godfrey:1985xj} for example) which prove
successful in describing the states below the
open-charm thresholds.  The difficulties aroused theorists' attention on
the coupled-channel
effects,
tetraquarks, mesonic molecules, quark-gluon hybrids, and  other
interpretations (see, e.g.,
\cite{Brambilla:2010cs} and the references therein).  The common consensus in
 these approaches is that many of these states can hardly be
accommodated in a conventional quarkonium description.

Among these $XYZ$ states, the $X(3915)$ was reported by Belle
 as a narrow resonance in the two-photon fusion process $\gamma\gamma\rightarrow
J/\psi\omega$~\cite{Uehara:2009tx}, in which both assignments of
$J^{PC}=0^{++}$ and $2^{++}$ are acceptable.   Soon after, this state
was
suggested to be the $\chi_{c0}(2P)$ state
in Ref.~\cite{Liu:2009fe}. {\it BABAR} confirmed the existence of the
$X(3915)$ and also suggested that its $J^{PC}$ is $0^{++}$ by studying the
angular distributions among the final leptons and pions of $J/\psi$ and
$\omega$~\cite{Lees:2012xs}. In the Particle Data Group (PDG) table~\cite{Agashe:2014kda}, this state is
quoted as $\chi_{c0}(2P)$ now. However, this assignment is questioned in
Ref.~\cite{Guo:2012tv} for the reason
that the properties of $X(3915)$ are far beyond the common
expectations to $\chi_{c0}(2P)$.
They suggested that the broad structure around 3.83GeV
, treated as backgrounds by Belle and {\it BABAR}, is contributed by
a $\chi_{c0}(2P)$ candidate,
whose mass coincides with the
predictions of the coupled-channel
models~\cite{Pennington:2007xr,Zhou:2013ada,Danilkin:2010cc}. The
$\chi_{c0}(2P)$ assignment to $X(3915)$ is also challenged by
Ref.\cite{Olsen:2015wna}, in which the author pointed out that this
assignment implies a confliction between the branching fractions
${\cal B}[\chi_{c0}(2P)\rightarrow J/\psi\omega]$ obtained from $\gamma\gamma\to
J/\psi \omega$ and $B\to K\,X(3915)$.
Moreover, the dominant decay mode of $\chi_{c0}(2P)$ is
 $\chi_{c0}(2P)\to D\bar D$, which is the Okubo-Zweig-Iizuka
allowed channel; therefore, some signal of $X(3915)$ in this channel
is expected.
There does exist a peak around $3.93$GeV, dubbed the $X(3930)$, in the
mass distributions of
the $\gamma\gamma\rightarrow D\bar{D}$ process from both Belle and {\it BABAR},
but the assignment of $2^{++}$ quantum
numbers to the resonance is consistent with the
data~\cite{Uehara:2005qd,Aubert:2010ab}.
By a closer examination of {\it BABAR}'s analysis against the assignment of
$2^{++}$ to $X(3915)$, one finds that the argument is
based on the helicity-2 dominance assumption which
originally comes from the quark model calculations on
the decay of a quarkonium  to two massless vector
particles~\cite{Krammer:1977an,PhysRevD.43.2161}.   However, if
the $X(3915)$ is not composed solely of quarkonium and  has
non-$q\bar q$ components as discussed in
\cite{Pennington:2007xr,Zhou:2013ada,Danilkin:2010cc}, this assumption
may fail and the helicity-0 contribution may also be sizable. In fact,
whether the helicity-2 contribution is dominant or not should be
determined by the experiment.

In this Letter, assuming a broad
$0^{++}$ resonance around $3.83$GeV and a narrow $2^{++}$ resonance
around $3.93$GeV, we first examine
the $\gamma\gamma\rightarrow D\bar{D}$ mass and angular distributions
to see whether these data are stringent enough to determine that the helicity-2
dominance assumption is indispensable, and we find that the  answer is
negative. Then by abandoning this assumption and assuming the same
$2^{++}$ resonance around $3.93$GeV in both $\gamma\gamma\rightarrow
D\bar{D}$ and 
$J/\psi\omega$ processes, we incorporate in our
analysis also the angular
distribution data of the $\gamma\gamma\rightarrow
J/\psi\omega$ process from {\it BABAR} which they used in determining the quantum
number of $X(3915)$. We find that the assignment of $2^{++}$ to
$X(3915)$ is
more consistent with the data than {\it BABAR}'s original assignment.
Furthermore, our results  also demonstrate that there is a large
helicity-0 contribution from the tensor resonance to the amplitude.
Regarding $X(3915)$ and $X(3930)$ to be the same state and
abandoning the helicity-2 dominance for this state are important for
the further theoretical and experimental studies on this state and are
helpful in understanding the properties of the mysterious $XYZ$
quarkonium-like states.

We first introduce our theoretical framework.
The differential cross section of $\gamma\gamma\rightarrow D\bar{D}$
could be represented by two independent helicity amplitudes
$\mathcal{M}_{+\pm}$ as
\bqa
\frac{d\sigma}{d\Omega}=\frac{1}{64\pi^2\rho(s)s}(|\mathcal{M}_{++}|^2+|\mathcal{M}_{+-}|^2),
\eqa
where $\rho(s)=\sqrt{(s-4m_D^2)/s}$. The partial-wave expansions of
$\mathcal{M}_{+\pm}$ are \cite{Collins:1977jy}
\bqa
\mathcal{M}_{++}(s,\mathrm{cos}\theta)=16\pi\sum_{J\geq 0}(2J+1) F_{J0}(s) d^J_{0,0}(\mathrm{cos}\theta),\nonumber\\
\mathcal{M}_{+-}(s,\mathrm{cos}\theta)=16\pi\sum_{J\geq 2}(2J+1)
F_{J2}(s) d^J_{2,0}(\mathrm{cos}\theta),
\label{eq:helicity-M}
\eqa
 in which  $F_{J,0}$ and $F_{J,2}$ are the partial-wave
amplitudes for helicity-0 and helicity-2 with vanishing odd-$J$ partial
waves.
 The $d$ functions are Wigner $d$-functions. So the cross section for
a certain partial wave with helicity $\lambda$ is
$\sigma_{J\lambda}=\frac{16\pi(2J+1)}{(\rho(s)s)}|F_{J\lambda}(s)|^2$.
 As a result of the coupled
channel unitarity~\cite{PhysRevD.35.1633}, when the energy region reaches
above the $J/\psi\omega$ threshold, the partial-wave amplitudes could
be expressed as
\bqa\label{coupleamp}
F_{J\lambda}(\gamma\gamma\rightarrow D\bar{D};s)&=&\alpha_{1;J\lambda}(s)T_{J}(D\bar{D}\rightarrow D\bar{D};s)\nonumber\\
&+&\alpha_{2;J\lambda}(s)T_{J}( J/\psi\omega\rightarrow D\bar{D};s),\nonumber\\
F_{J\lambda}(\gamma\gamma\rightarrow J/\psi\omega;s)&=&\alpha_{1;J\lambda}(s)T_{J}(D\bar{D}\rightarrow J/\psi\omega;s)\nonumber\\
&&\hspace{-1cm}+\alpha_{2;J\lambda}(s)T_{J}(J/\psi\omega\rightarrow J/\psi\omega;s),
\eqa
  where $T_J$ is the corresponding hadronic partial-wave amplitude.  The coupling functions, $\alpha_{i;J\lambda}$, being smooth real
functions in the physical region,  only contain the left-hand 
cut contributions~\cite{Dai:2014zta,PhysRevD.35.1633}.
Under the pole dominance assumption, if the pole's couplings to the
two channels are parameterized by $g_{D\bar{D}}$ and $g_{J/\psi\omega}$, respectively, one can easily find
that the ratios $F_{J0}/F_{J2}$ in the
two channels are equal at the pole position.  Because of  
the lack of enough data, we could not make a
close-to-model-independent analysis as the discussion of
$\gamma\gamma\rightarrow\pi\pi,K\bar{K}$ in
Refs.~\cite{Morgan:1987wi,Dai:2014zta} but make an analysis
phenomenologically.
If a certain $L$-th partial-wave amplitude is mainly contributed
through a resonance, one can just write the hadronic scattering
amplitude in an elastic relativistic Breit-Wigner form as used in
experimental analyses~\footnote{In principle, a coupled-channel
Breit-Wigner form is required to make a combined analysis for two
channels. However, to compare with the experimental analyses, which
analyze different channels separately, an elastic Breit-Wigner form is
good enough to illustrate our point.}:
\bqa
T_L(s)=\frac{M\Gamma(s)}{M^2-s-iM\Gamma(s)},\,\,
\Gamma(s)=\Gamma\left(\frac{p}{p_0}\right)^{2L+1}\frac{M}{\sqrt{s}}F_L^2(s),
\nonumber\eqa
with $M$ the nominal mass of the resonance and $\Gamma$ its total
width. The three-momentum of an outgoing $D$ meson in the $D\bar{D}$
center of mass frame is denoted by $p$, and $p_0$ is the corresponding value for $\sqrt{s}=M$. The Blatt-Weisskopf factor
$F_0=1$
and $F_2=\frac{\sqrt{9+3(p_0R)^2+(p_0R)^4}}{\sqrt{9+3(pR)^2+(pR)^4}}
$,
with $R=1.5\,\mathrm{GeV}^{-1}$ as used in Ref.~\cite{Aubert:2010ab}.

We assume that the lowest two partial waves, the S wave and D wave,
are contributed by a $0^{++}$ resonance and a $2^{++}$
one, respectively, dominating the $\gamma\gamma\rightarrow D\bar{D}$
process below $4.2\, \mathrm{GeV}$. The $0^{++}$ resonance contributes
through the S wave of helicity-0 amplitude, while the $2^{++}$
resonance contributes to both helicity-0 and helicity-2 amplitudes. For
simplicity, we parameterize every $\alpha_{i,J\lambda}(s)$ function
by one parameter instead of by nonsingular polynomials with more free
parameters, and for the lack of information about the $D\bar{D}\rightarrow J/\psi\omega$ scattering amplitude, the relative strength and phase between the S wave and D wave of helicity-0 amplitude are parametrized by a complex number as $\beta e^{i\phi}$.

Thus,  the helicity amplitudes of $\gamma\gamma\rightarrow D\bar{D}$ are represented phenomenologically as
\bqa
\mathcal{M}_{++}&=&16\pi[\mathcal{A}_0(s)+\beta_1 e^{i\phi_1}\mathcal{A}_2(s)\times 5\times d^2_{0,0}(\mathrm{cos}\theta)],\nonumber\\
\mathcal{M}_{+-}&=&16\pi[\beta_2 e^{i\phi_2}\mathcal{B}_2(s)\times 5\times d^2_{2,0}(\mathrm{cos}\theta)],
\eqa
where
$
\mathcal{A}_0(s)=\frac{M_{\chi_{c0'}}\Gamma_{\chi_{c0'}}(s)}{M_{\chi_{c0'}}^2-s-iM_{\chi_{c0'}}\Gamma_{\chi_{c0'}}(s)}$
and 
$\mathcal{A}_2(s)=\mathcal{B}_2(s)=\frac{M_{\chi_{c2'}}\Gamma_{\chi_{c2'}}(s)}{M_{\chi_{c2'}}^2-s-iM_{\chi_{c2'}}\Gamma_{\chi_{c2'}}(s)}$.
One could use these amplitudes to fit the
$\gamma\gamma\rightarrow D\bar{D}$ mass distributions and the
angular distribution  simultaneously. The
parameters are $M_{\chi_{c0'}}$, $\Gamma_{\chi_{c0'}}$,
$M_{\chi_{c2'}}$, $\Gamma_{\chi_{c2'}}$, $\beta_1$, $\phi_1$, and
$\beta_2$. The phase $\phi_2$ will not be  determined in the fit, and
is not important in our discussion indeed. There are other two
normalization parameters to rescale the mass and  angular distribution
events respectively.

\begin{table*}
\caption{\label{fit}The fitted parameters for
Belle~\cite{Uehara:2005qd} and {\it BABAR}~\cite{Aubert:2010ab} data.
}
\begin{tabular}{|c|c|c|c|c|c|c|c|c|}

  \hline
   Parameters & ``Fit Belle 1" & ``Fit Belle 2" & ``Fit Belle 3" & &``Fit {\it BABAR} 1" & ``Fit {\it BABAR} 2" & ``Fit {\it BABAR} 3"  \\
\hline
  $\chi^2/{\mathrm d.o.f}$  &44.8/(47+10$-$9)  & 45.2/(47+10$-$7) &
55.5/(47+10$-$8)& &71.9/(47+10$-$9) & 73.7/(47+10$-$7) &
73.1/(47+10$-$8) \\
\hline
  $M_{\chi_{c0'}}(\mathrm{\mathrm{GeV}})$  & $3.817\pm 0.009$ & $3.814\pm 0.006$ & $3.820\pm 0.009$ & &$3.853\pm 0.009$ & $3.851\pm0.009$ & $3.853\pm 0.009$\\
\hline
  $\Gamma_{\chi_{c0'}}(\mathrm{\mathrm{GeV}})$  &  $0.163\pm 0.033$& $0.155\pm 0.020$ & $0.201\pm 0.019$ &  & $0.229\pm0.031$ & $0.227\pm 0.032$ &$0.233\pm 0.030$\\
\hline
  $M_{\chi_{c2'}}(\mathrm{\mathrm{GeV}})$  & $3.925\pm 0.003$ & $3.925\pm 0.005$ & $3.924\pm 0.009$ & &$3.932\pm 0.001$ &$3.932\pm0.001$ &$3.932\pm 0.001$\\
\hline
  $\Gamma_{\chi_{c2'}}(\mathrm{\mathrm{GeV}})$  & $0.035\pm 0.005$ & $0.036\pm 0.005$ & $0.031\pm 0.005$& &$0.021\pm 0.004$ &$0.021\pm 0.005$ &$0.020\pm 0.004$ \\
\hline
  $\beta_1$  & $0.147\pm 0.201$ & $0 $& $0.5$& &$0.290\pm 0.237$ & $0$& $0.5$ \\
\hline
  $\phi_1(\mathrm{rad})$  & $2.850\pm 0.513$ &  & $3.653\pm 0.389$&  & $3.713\pm 1.326$& & $3.700\pm 0.597$ \\
\hline
  $\beta_2$  & $0.559\pm 0.077$ & $0.586\pm0.051$ & $0.388\pm 0.086$ &
&$0.514\pm 0.151$ & $0.599 \pm0.056$& $0.330\pm 0.101$\\

  \hline
  \end{tabular}
\end{table*}

\begin{figure}[t]%
\begin{center}%
\includegraphics[height=32mm]{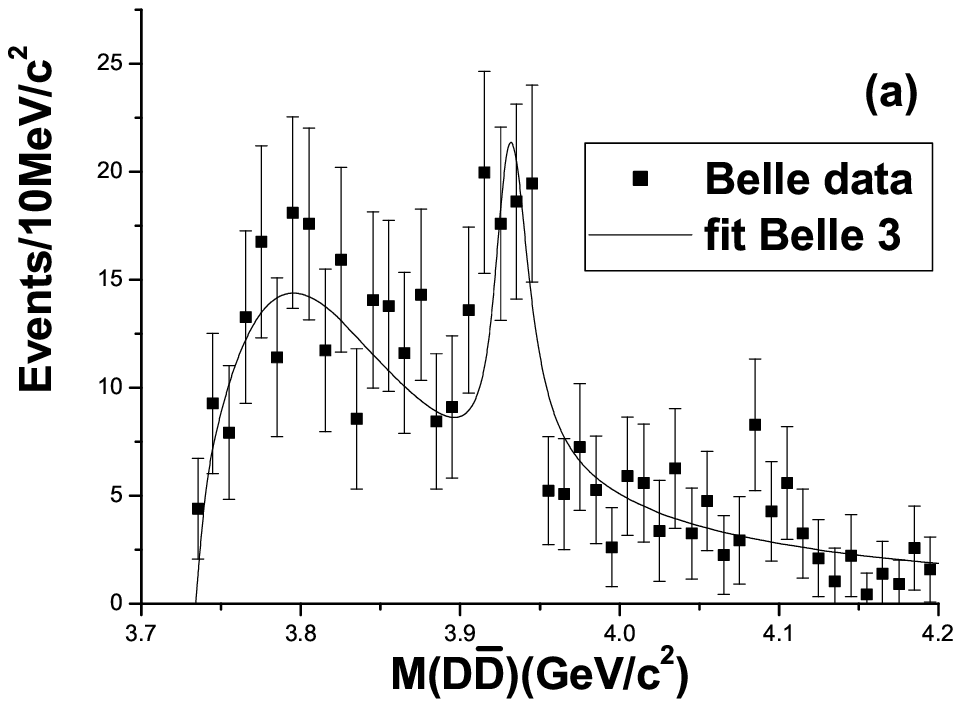}
\includegraphics[height=32mm]{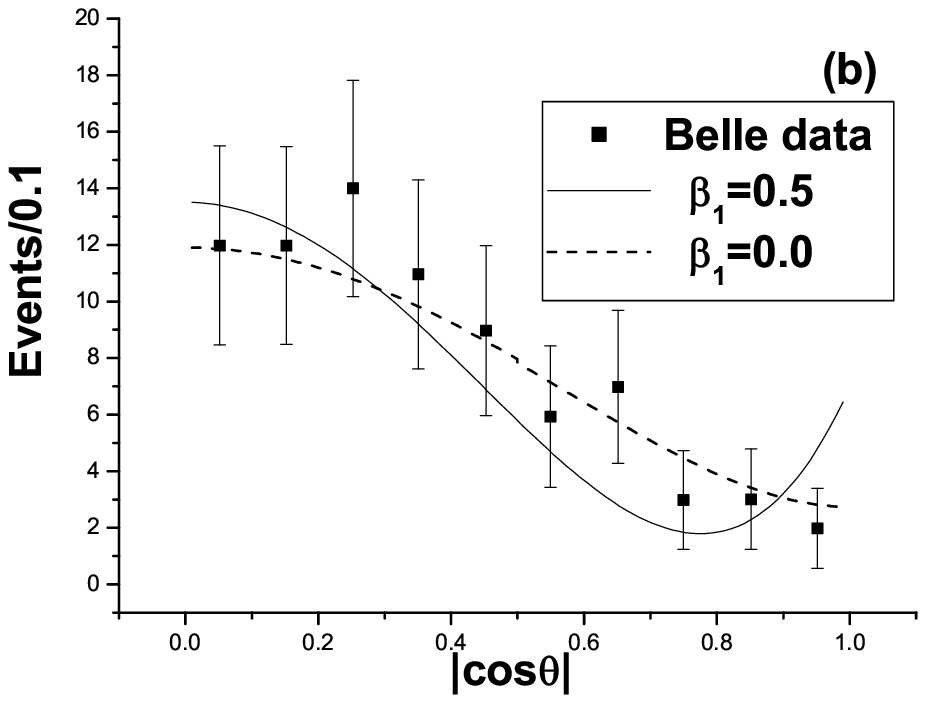}
\includegraphics[height=32mm]{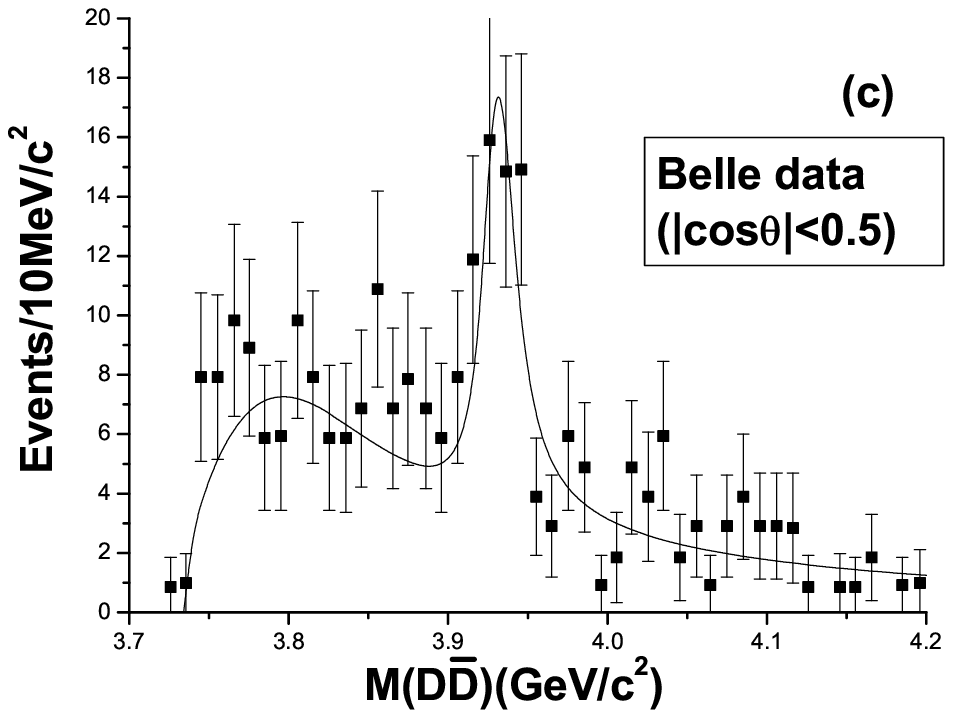}
\includegraphics[height=32mm]{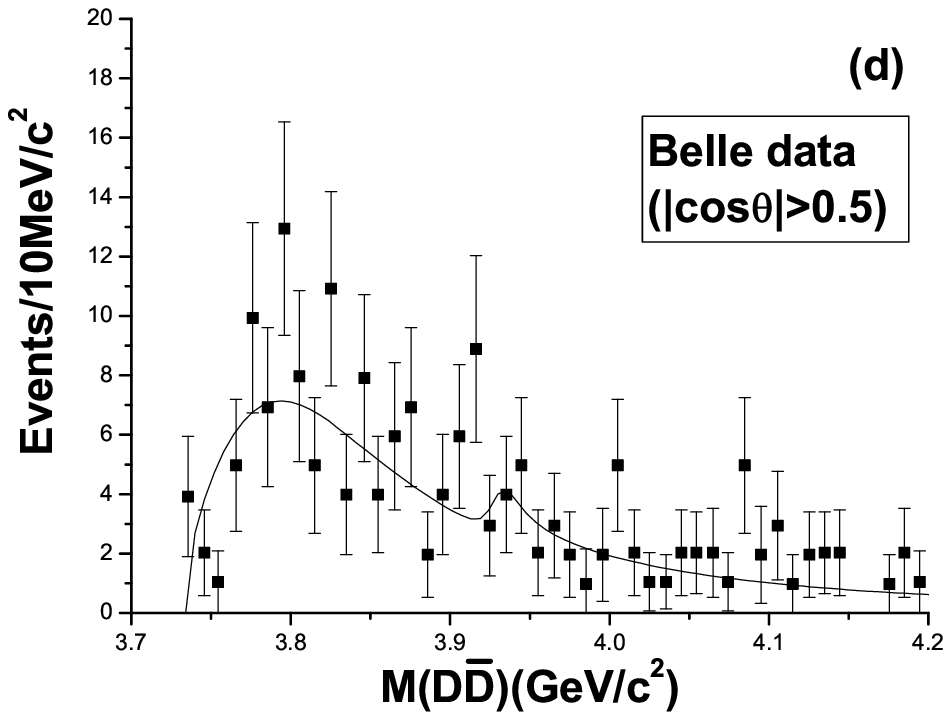}
\caption{\label{belle} (a)The $M(D\bar{D})$ distribution in
$\gamma\gamma\rightarrow D\bar{D}$ of Belle data compared with the
result of fit Belle 3. (b) The angular distributions of Belle data
compared with fit Belle 2 and fit Belle 3. (c) and (d) are the typical
reproduced mass distributions for $|\cos\theta|<0.5$ and
$|\cos\theta|>0.5$, respectively, for three fits.}
\end{center}%
\end{figure}

As a test of our formulas and assumptions, we made a fit to the mass
distribution data below $4.2\,\mathrm{GeV}$ and angular distribution
data in the range of $3.91\,\mathrm{GeV}<\sqrt{s}<3.95\,\mathrm{GeV}$ from
Belle and {\it BABAR}, respectively,  with all
parameters free. The fit results are shown in the fit Belle 1 and fit {\it BABAR} 1
columns of the corresponding data sets in Table~\ref{fit}. The parameters
of the $J=2$ narrow resonance are close to the values presented by
Belle~\cite{Uehara:2005qd} and {\it BABAR}~\cite{Aubert:2010ab}, while those of the broad $J=0$ resonance are similar
to the values in Ref.~\cite{Guo:2012tv}.
Although we fit the total mass distributions, the separate mass
distributions for $|\cos \theta |>0.5$ and $|\cos \theta |<0.5$ can also be
reproduced automatically as shown in Figs.~\ref{belle}(c) and \ref{belle}(d), which also demonstrates the reasonability of
our method.

\begin{figure}[t]%
\begin{center}%
\includegraphics[height=32mm]{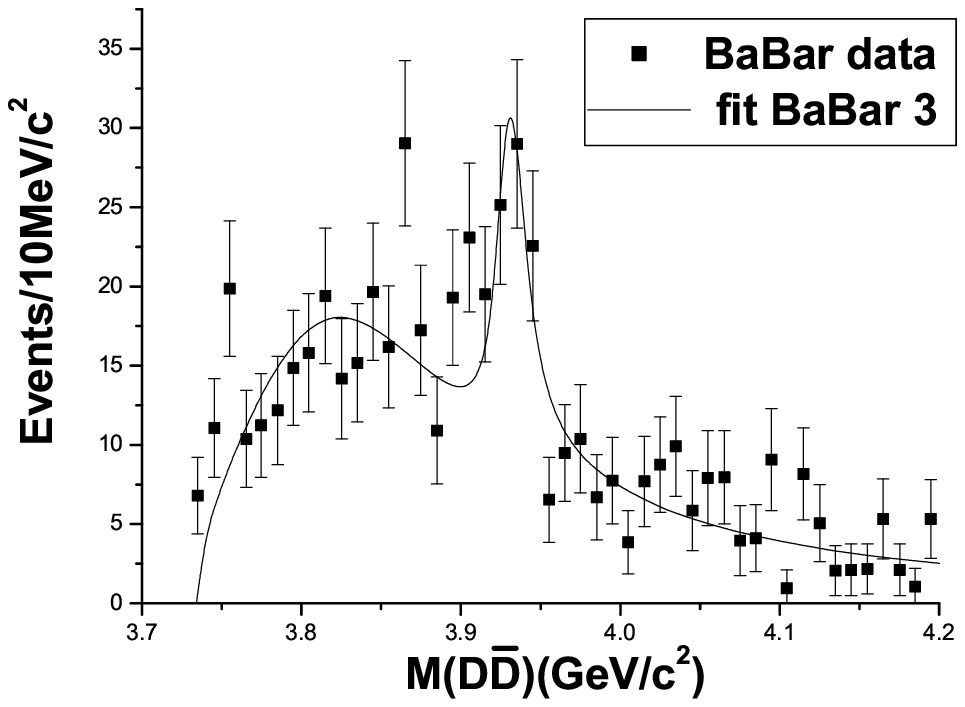}
\includegraphics[height=32mm]{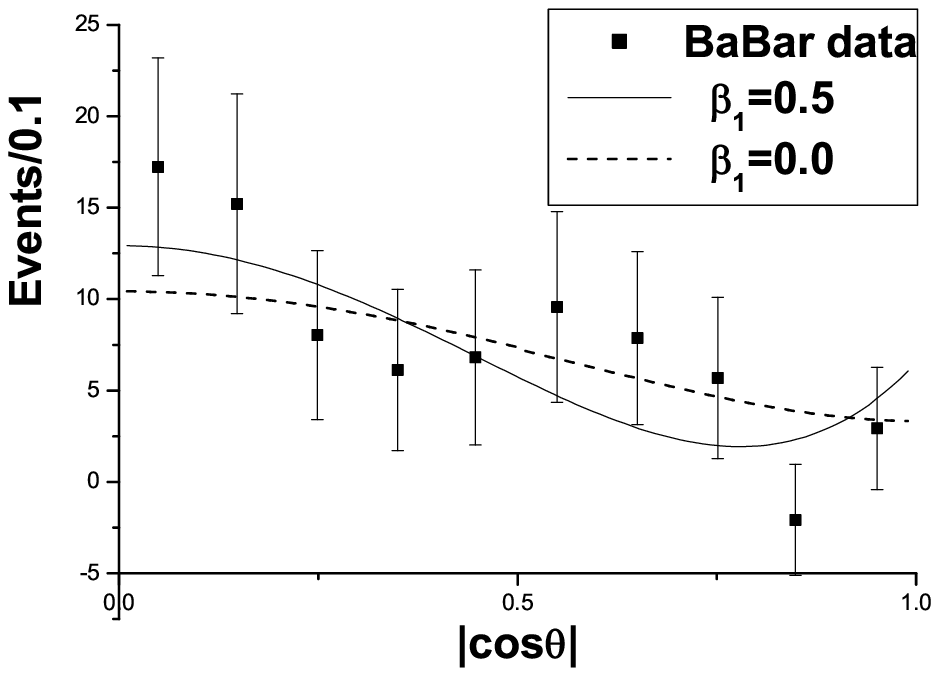}
\caption{\label{babar}(a)The $M(D\bar{D})$ distribution in
$\gamma\gamma\rightarrow D\bar{D}$ of \textsl{BABAR} data compared
with the result of fit \textsl{BABAR} 3. (b)The angular distribution
of \textsl{BABAR} data compared with the results of fit \textsl{BABAR}
2 and fit \textsl{BABAR} 3.
}
\end{center}%
\end{figure}

The ratio $\mathcal{R}=(\beta_1/\beta_2)^2$ denotes
the relative strength between the
helicity-0  and the helicity-2 contributions from the tensor
resonance. In both fit results, $\beta_1$ parameters have large errors,
which means that the experimental data of $\gamma\gamma\to D\bar D$
 do not impose a strong constraint on the helicity-0
component from the tensor resonance and may allow for a
large helicity-0 contribution  comparable to or even larger than the
helicity-2 contribution. In order to test the necessity of the
helicity-2 dominance, we perform another
two fits to each data set, by fixing $\beta_1=0$ and $0.5$, respectively.
Both fit results, shown in Table.\ref{fit}, present acceptable qualities as
shown in Figs.~\ref{belle} and \ref{babar}.
Although the $\chi^2/{\rm d.o.f}$  for fit
Belle 3 is $1.12$, which deviates a little farther from $1$
than $0.93$ for the all-free fit, it is still acceptable.
The fit {\it BABAR} 3 result
gives almost the same $\chi^2/{\rm d.o.f.}$ as the one for the all-free fit.
Anyway, all these fits demonstrate that the helicity-2
dominance assumption is not necessary in determining  the $X(3930)$.
So, we will abandon this assumption in the following discussion.

Without the helicity-2 dominance assumption, we can reexamine the
quantum numbers of the $X(3915)$ by incorporating the data for
$X\rightarrow J/\psi(\rightarrow l^+
l^-)\omega(\rightarrow\pi^+\pi^-\pi^0)$ from {\it BABAR}. Because of  the
degeneracy of the $X(3915)$ and the $X(3930)$, we assume that the
$X(3915)$ and the $X(3930)$ are the same tensor state and check whether
it is consistent with the angular distribution data used by {\it BABAR} in
determining the quantum numbers of the $X(3915)$.

Let us recall some basic
definitions and results from Ref.~\cite{PhysRevD.70.094023}.
$\theta^*_l$ is defined as the angle between the momentum of
the positively charged lepton from $J/\psi$ decay ($l^+$) and the
$\gamma\gamma$ axis in the $J/\psi\omega$ rest frame. $\theta^*_n$ is
defined as the angle between the normal to the decay plane of the
$\omega$ (defined as $\vec{n}$) and the
$\gamma\gamma$ axis in the $J/\psi\omega$ rest frame. $\theta_{ln}$ is the angle
between the lepton $l^+$ from $J/\psi$ decay and the normal of
the $\omega$ decay plane, $\vec{n}$~\cite{Lees:2012xs}.
The $\theta^*_l$ and $\theta^*_n$ angular distributions in
$X\rightarrow J/\psi(\rightarrow l^+l^-)\omega(\rightarrow
\pi^+\pi^-\pi^0)$ for a $2^{++}$ $X$ state produced in a $\gamma\gamma$ collision read
\bqa
\frac{d\Gamma}{d\mathrm{cos}\theta^*_l}&=&\frac{1}{8}[6|A_{\pm 2}|^2(1+\mathrm{cos}^2\theta^*_l)+|A_0|^2(5-3\mathrm{cos}^2\theta^*_l)],\nonumber\\
\frac{d\Gamma}{d\mathrm{cos}\theta^*_n}&=&\frac{1}{4}[6|A_{\pm
2}|^2(1-\mathrm{cos}^2\theta^*_n)+|A_0|^2(1+3\mathrm{cos}^2\theta^*_n)],\nonumber
\eqa
with the helicity amplitudes $A_0$ and $A_{\pm 2}$  suitably normalized,
while the $\theta_{ln}$ angular distribution for the decay of a
$2^{++}$ $X$ state produced in a $\gamma\gamma$ collision is
$\frac{1}{\Gamma}\frac{d\Gamma}{d\mathrm{cos}\theta_{ln}}=\frac{3}{40}[7-\cos^2\theta_{ln}]$.
The angular distributions of lepton or pion for a $0^{++}$ state
do not show any dependence on the angles of $\theta^*_l$ or
$\theta^*_n$, and the angular dependence for a $0^{++}$ state on $\theta_{ln}$ is
$\frac{1}{\Gamma}\frac{d\Gamma}{d\mathrm{cos}\theta_{ln}}=\frac{3}{4}\mathrm{sin}^2\theta_{ln}.
$

\begin{figure}[t]%
\begin{center}%
\includegraphics[height=32mm]{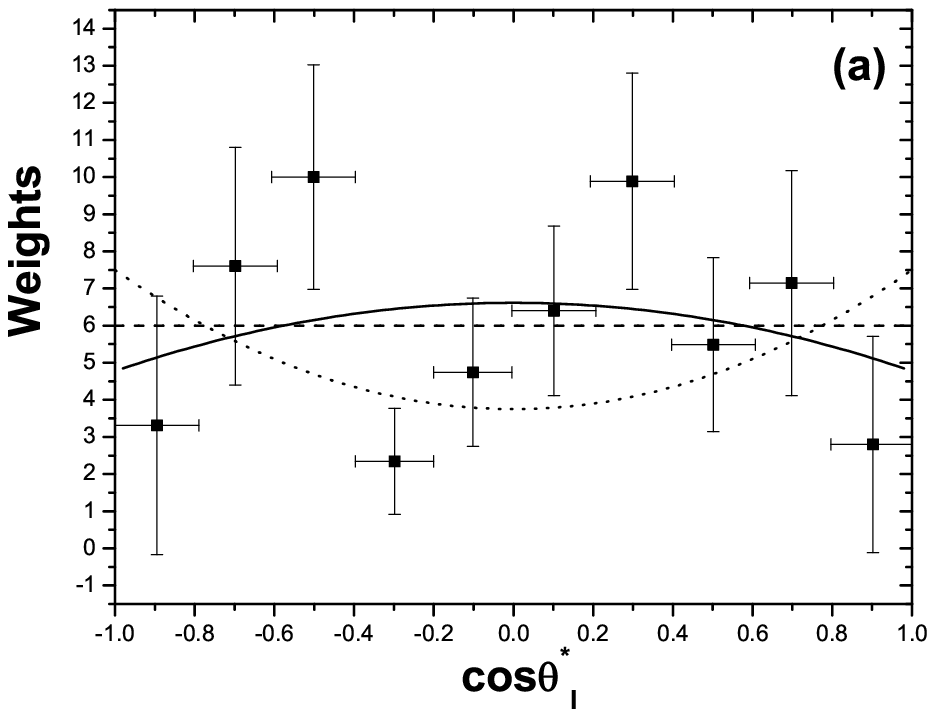}
\includegraphics[height=32mm]{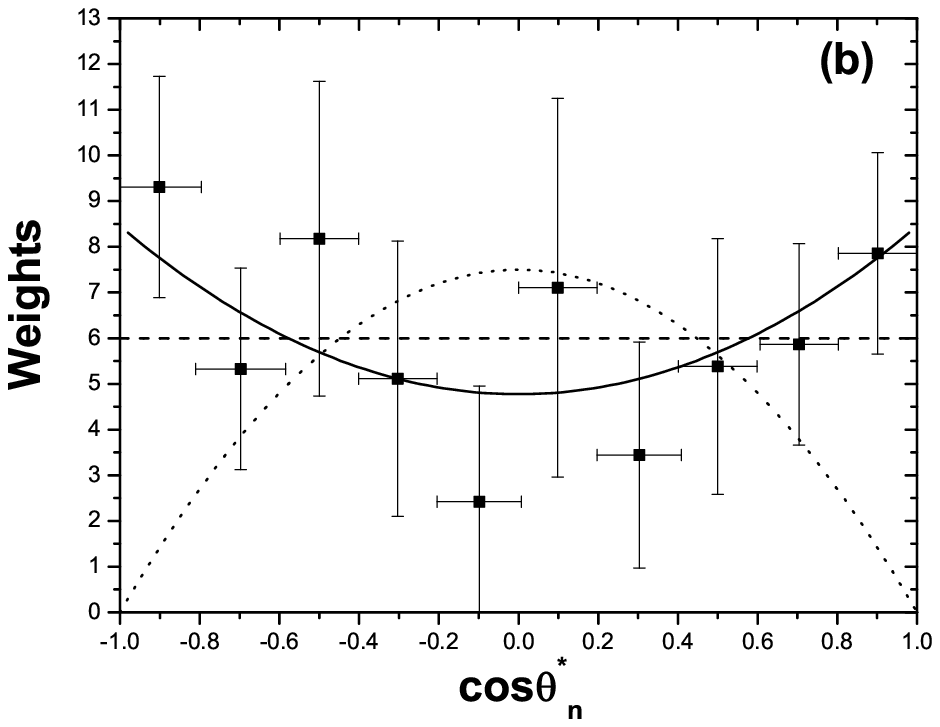}
\includegraphics[height=32mm]{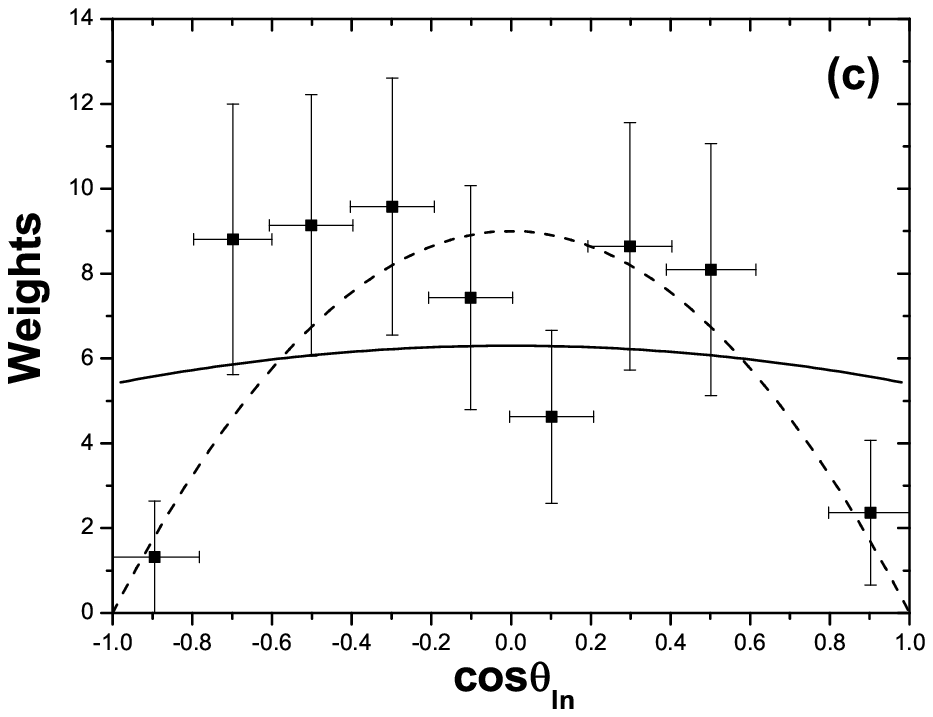}
\caption{\label{ang3915} 
 The angular distributions for (a) $\mathrm{cos}\theta^*_l$, (b)
$\mathrm{cos}\theta^*_n$, and (c) $\mathrm{cos}\theta_{ln}$ of $X(3915)$. Dashed line: intermediate $0^{++}$
resonance; solid line:  $2^{++}$ resonance in the combined
fit; dotted line : $2^{++}$ resonance
with helicity-2 dominance used by \textsl{BABAR} in
Ref.~\cite{Lees:2012xs}.
}
\end{center}%
\end{figure}

 The ratios ${\cal R}$
of the $2^{++}$ intermediate state in $\gamma\gamma\rightarrow
J/\psi \omega, D\bar{D}$ are the same
according to Eq.(\ref{coupleamp}) and the pole dominance assumption.
We then perform another fit combining the previous $\gamma\gamma\to
D\bar D$ data and the above three angular-distribution data from
$\gamma\gamma\to J/\psi\omega$, setting all the parameters free.
Since Belle has no angular-distribution data for
$\gamma\gamma\to J/\psi\omega$ channel, we use only {\it BABAR}'s data for
both channels for consistency.
In the combined fit, the $\chi^2$ contributed by $\gamma\gamma\to D\bar D$ data is
$74.7$ which is almost the same as the one for fit {\it BABAR} 1.
The $\chi^2$ for the $\cos \theta_l^*$ distribution in
Fig.~\ref{ang3915}(a) is improved from 11.2 for $0^{++}$ assignment
to 10.1 for $2^{++}$ assignment and the one for 
$\cos \theta_n^*$ distribution in Fig.~\ref{ang3915}(b) is improved from 6.9 to 4.5.
In  {\it BABAR}'s argument, it is just because of the much better $\chi^2$
 of the dashed line than the dotted line in the second graph that they strongly
suggest the $0^{++}$ assignment to $X(3915)$. So, their reasoning is
not convincing any more.
The angular distribution of the $\cos\theta_{ln}$ does not depend on the
helicity since it averages different moving directions of $X(3915)$. The curves in Fig.~\ref{ang3915}(c) are not fits but predictions. 
However, the analysis based on these data for $\cos\theta_{ln}$ 
distribution is not quite reliable. In principle, the distribution
must be symmetric with respect to $\cos\theta_{ln}=0$, but in the
experimental data in the region $0.6<\cos \theta_{ln}<0.8$ there is
no events,  while in the
mirror region,  $-0.8<\cos \theta_{ln}<-0.6$, there are around nine 
events detected. This clearly demonstrates that the statistics is still too low
to produce a good $\cos\theta_{ln}$ distribution. Furthermore, the $\chi^2$ also strongly depends on the missing
events in Fig.~\ref{ang3915}(c).
In {\it BABAR}'s analysis, by setting the data in the region of $0.6<\cos
\theta_{ln}<0.8$ to be zero, they obtained that the $\chi^2$ for $\cos \theta_{ln}$ in Fig.~\ref{ang3915}(c) is 18.0 for the $2^{++}$ assignment,
worse than 12.5 for the $0^{++}$ assignment. However, as a reasonable
assumption, if the data in this region are the same as that of
$-0.8<\cos \theta_{ln}<-0.6$ due to the symmetry, the $\chi^2$ of $\cos
\theta_{ln}$ distribution will be significantly reduced to 10.7 for
$2^{++}$ vs 9.2 for $0^{++}$. Because of the apparent poor quality of the $\cos
\theta_{ln}$ distribution data, they cannot be used to distinguish which
assignment is better. Thus, the total $\chi^2/{\rm d.o.f}$ in our analysis
improves from $1.50$ for the fit {\it BABAR} 1 to about $1.27$, which
demonstrates that the experimental data favor the $2^{++}$ assignment
of $X(3915)$.

With the
tensor assignment, the $X(3915)$   could
be the same state as the $X(3930)$ due to the degeneracies of their
masses and widths. Since the global fit automatically produces $\beta_1:\beta_2\simeq 0.48:0.31$ for the
tensor assignment, our result is sufficient to illustrate that,
the experimental data prefer a sizable helicity-0
contribution which may even be larger than the helicity-2
contribution.
As stated before, the helicity-2 dominance is a consequence of
assuming  that $X(3915)/X(3930)$ is a quarkonium
state~\cite{Krammer:1977an,PhysRevD.43.2161}. The large helicity-0
contribution in our scenario means that $X(3915)/X(3930)$ might not be
a pure $q\bar{q}$ quarkonium state, but with a significant
non-$q\bar{q}$ component.  Actually, most charmoniumlike states above
the open-charm threshold can hardly be understood as conventional
quarkonium states, but can be described well in the spirit of coupled-channel
effects~\cite{Eichten:1978tg,Eichten:1979ms,Heikkila:1983wd,vanBeveren:1983td,Pennington:2007xr,Zhou:2013ada,Danilkin:2010cc}
or as  molecular states~\cite{Liang:2009sp}, in
which the states contain a significant portion of non-$q\bar q$
components.

In conclusion, in this Letter, we present that the
$\gamma\gamma\rightarrow D\bar{D},J/\psi\omega$ data favor a
possibility that the $X(3915)$, listed in the PDG table as the 
$\chi_{c0}(2P)$ state, is just the $2^{++}$ state $X(3930)$. The data
also prefer that this tensor state has a sizable helicity-0
contribution in the $\gamma\gamma\rightarrow D\bar{D}, J/\psi\omega$
amplitudes, comparable to or maybe larger than the helicity-2
contribution. This is a hint of a sizable non-$q\bar q$ component in the
state.    Our results of
identifying the $X(3915)$ with the $X(3930)$ and abandoning the
helicity-2 dominance may inspire further theoretical and experimental
explorations on the properties of the charmoniumlike states.

\begin{acknowledgments}
Helpful discussions with Yan-Wen Liu are
appreciated.  We also thank Antimo Palano for providing the
information of the data. This work is supported by the National
Natural Science Foundation of China under Grants No.11375044, No. 11105138, and
No. 11235010. Z.Z. thanks the Project Sponsored by the Scientific
Research Foundation for the Returned Overseas Chinese Scholars, State
Education Ministry. Z.X. is also partly supported by the Fundamental Research Funds for the
Central Universities under Grant No.WK2030040020.
\end{acknowledgments}

\bibliographystyle{apsrev4-1}
\bibliography{chi02p}

\end{document}